\documentstyle[sprocl]{article}

\input{epsf}

\def\Lam{\Lambda}  \def\Sig{\Sigma}
\def\wave{\simeq}
\def\ie{{\it i.e.}}
\def\di{{\Delta I}}

\def\vsig{\mbox{\boldmath$\sigma$}}
\let\vec\bf
\def\SU#1{SU(#1)}
\def\Lam{\Lambda}  \def\Sig{\Sigma}
\def\wave{\simeq}
\def\ie{{\it i.e.}}
\def\di{{\Delta I}}
\def\pip{{\pi^+}}  \def\pim{{\pi^-}}

\begin{document}

\title{Roles of Quarks in Strong and Weak YN Interactions}
\author{M.~Oka, Y.~Tani, T.~Inoue$^{(a)}$ and K.~Sasaki}
\address{Department of Physics,
        Tokyo Institute of Technology \\
        Meguro, Tokyo, 152-8551 JAPAN\\
        E-mail: oka@th.phys.titech.ac.jp}

\address{$^{(a)}$Department of Physics, Univ.\ of Tokyo\\
         Hongo, Tokyo, 113-0033 JAPAN}

\maketitle
\abstracts{
 The short range parts of the strong and weak hyperon-nucleon
 interactions are studied with quark substructure of the baryons taken
 into account.
 A summary of the quark cluster model calculations of the baryon-baryon
 interactions is presented.  It is pointed out that the spin-flavor
 symmetry structure determines the qualitative behaviors of the short-range
 YN interactions.
 We also show that the spin-orbit YN forces give characteristic
 behaviors, which are distinct from the meson exchange forces.
 Weak decays of $\Lam$ in hypernuclei are described by the direct
 quark mechanism.  We find that the neutron induced nonmesonic decay
 is enhanced due to the direct quark transition. Importance of the
 $\di=3/2$ decay amplitudes is emphasized both for the nonmesonic weak
 decays and the $\pip$ emission in the $\Lam$ decay.}

\section{Introduction}

Recent experimental activities in hypernuclear physics provide us
with high quality data of production, spectra and decays of hypernuclei.
The accumulation of such data accelerates quantitative analyses
of the strong and weak interactions of hyperons.
The hyperon brings a new flavor, strangeness, to the up-down world of
nuclei.  Several interesting questions, such as the role of the Pauli
principle for the hyperon, possible existence of dibaryon resonances,
etc., are yet to be answered.
It is therefore desirable to understand the behaviors of the hyperon
in nuclear environment in the context of QCD.
We here consider the quark substructure of the baryons and study the
roles of the explicit quark degrees of freedom.
The spin and flavor symmetry of quarks plays the most important
role as the SU(3) symmetry was the driving force for developing the
quark model of the hadrons.  We mostly concentrate on general
arguments that are based only on the symmetry structure, and for a
numerical calculation we introduce a model, where the simple SU(3)
constituent quark model and one-gluon exchange interactions as well as
the confining force among the quarks are employed.

The weak interaction is another interesting subject, which can be
studied in detail in the strangeness system.
Hypernuclei are rich sources of weak strangeness decay in nuclear
medium, where a new decay mode, $\Lam N\to NN$ is realized.
We point out that the $\di=1/2$ rule is not necessarily satisfied for
nonmesonic decays of hyperons, and that the neutron induced
nonmesonic decay is strongly enhanced due to the direct quark transition.
It is also found that $\pip$ decay of hypernuclei is generated mainly
by the $\di=3/2$ weak transitions.

In sect.\ 2, we present the quark model description of the baryon-baryon
interaction including hyperons.{\cite{YITP}}
The antisymmetric spin-orbit force is examined in sect.\ 3 from the \SU3\
flavor symmetry point of view.{\cite{TO}}
In sect.\ 4, we introduce the direct quark mechanism for nonmesonic weak
decays of $\Lam$.{\cite{ITO,IOMI,SIO}}
In sect.\ 5, the $\pip$ decay of hypernuclei and the role of the
$\di=3/2$ weak transitions are discussed.{\cite{piplus}}

\section{Short range part of YN Interactions}

We first summarize the main results of the quark model study of the
hyperon-nucleon (YN) interactions.  Although the details depend on
models of the quark dynamics we choose, some robust qualitative
features are derived from the symmetry argument.

In studying the YN interactions, it is natural to follow
descriptions of the nuclear force.  The
long-range part of the nuclear force is explained very well in terms
of one-pion exchange mechanism, while heavy
mesons as well as
multi-pion exchanges are necessary for the medium range
part.{\cite{Nijmegen,DF,Bonn,Julich}
One-boson exchange potential models, in which
two- (and multi-) pion exchanges are taken into account in terms of
$\sigma$ and $\rho$ exchanges, are fairly successful in accounting the
large amount of data for nucleon-nucleon scattering.
Yet the short-range part of the nuclear force is not fully
understood microscopically, that is, a repulsive core (hard or soft) is
often introduced phenomenologically to
explain the NN scattering phase shifts around $E\wave 100-200$ MeV in
the center of mass system.  Indeed, this is the region where the
internal quark-gluon structure of the nucleon must play important roles.

It was pointed out that the quark exchange force between two nucleons
gives strong repulsion at short distances.{\cite{OY}}
The exchange force
is induced by the quark antisymmetrization and therefore is nonlocal and
of short-range determined by the size of the quark content of the
nucleon.
The most important feature of the quark exchange force is its
dependence on the spin-flavor symmetry of two-baryon states.
A close analogy is found in the hydrogen molecule, where two
electrons orbit around two protons.  As the total spin of the
electrons specifies the symmetry of the spin wave function, the sign
of the exchange force is determined according to the spin.
The symmetric orbital state is allowed only for $S=0$, while the
exchange force is strongly repulsive for $S=1$.
Similar state dependencies appear in the quark exchange force, where
the spin-flavor \SU6\ symmetry determines the properties of the
exchange interactions.

We applied the quark cluster model description to the short-range
YN interactions.{\cite{OSY}}  We found that the flavor singlet
combination of $\Lam\Lam-N\Xi-\Sig\Sig$ has no repulsion
induced by the quark exchange.  As this state is known
to be favored by the magnetic part of the
one-gluon exchange interaction (color-magnetic interaction), a bound
or a resonance state called H dibaryon may exist.{\cite{Jaffe,hakone}}
On the other hand, most other YN and YY channels have strong
repulsion at short distances.

Another interesting observation made in ref.~\cite{OSY}\ is that the
$S$ wave $\Sig N$ interaction depends strongly on the total spin and
isospin.  The $S$-wave $\Sig N$ ($I=1/2$, $S=0$) and
$\Sig N$ ($I=3/2$, $S=1$)
states belong mainly to the [51] symmetric irreducible
representation of the spin-flavor \SU6\ symmetry.
This is the representation in which a
Pauli forbidden state appears in the $L=0$ orbital motion.  The Pauli
principle forbids two baryons to get closer and thus gives a strong
repulsion.  The other spin-isospin states do not belong to this
symmetry and therefore the short range repulsion is weaker.
This qualitative argument was confirmed in realistic quark cluster
model calculations of the YN interactions.{\cite{OSY}}
Recent analyses by Niigata-Kyoto group show that the strong repulsion
remains after combining the quark exchange interaction with the
long-range meson exchange attraction.{\cite{NSF}}
No experimental evidence is yet available to be able to
confirm the strong state dependencies.
More $\Sigma N$ scattering data are anticipated very much.

\section{Antisymmetric Spin-Orbit Forces}

One of the interesting features of the hyperon-nucleon interactions is the
properties of the spin-orbit force.
The Galilei invariant spin-orbit force consists of symmetric LS (SLS) and
antisymmetric LS (ALS) terms,
\begin{eqnarray}
   V_{SO} &=& V_{SLS}({\vsig_1}+{\vsig_2})\cdot {\vec L}
  +  V_{ALS}({\vsig_1}-{\vsig_2})\cdot {\vec L} \nonumber\\
   &=&  (V_{SLS}+  V_{ALS})\,{\vsig_1}\cdot {\vec L}
   +(V_{SLS}-  V_{ALS})\,{\vsig_2}\cdot {\vec L}
\end{eqnarray}
Because the ALS operator $(\vsig_1-\vsig_2)\cdot \vec L$
is antisymmetric with respect to the exchange of two baryons,
$V_{ALS}$ should be zero between like baryons.
In the nuclear force, ALS between proton and neutron breaks the isospin
symmetry and is classified as
a type IV charge symmetry breaking (CSB) force.
Evidence of such a CSB force was
given by measuring the difference between the proton and
neutron analyzing powers in the $n-p$ scattering experiments.
The results show that this force is very weak, supporting the isospin
invariance of the nuclear force.

On the contrary, the ALS forces in the hyperon-nucleon interactions do not
vanish even in the SU(3) flavor symmetric limit.
For hyperon-nucleon systems, ALS seems as strong as
the SLS part.
If the magnitudes of SLS and ALS are comparable, then
the single particle LS force for one of the baryons, (ex.\ $\Lambda$)
inside (hyper)nuclei is much weaker than that for the other baryon (nucleon).
Since recent experiment suggests that the single particle LS force for
$\Lambda$ might be sizable contrary to the wide belief of vanishing LS
force for $\Lambda$,  it is extremely important to  pin down the
magnitude of the two-body LS force.

\subsection{\SU3\ symmetry for ALS}
\def\rep#1{{\bf {#1}}}

First we study the properties of the YN ALS forces from the SU(3)
symmetry point of view.
For the octet baryons, the baryon-baryon interactions can be
classified in terms of the SU(3) irreducible representations given by
\begin{equation}
 \rep8 \times \rep8 = \rep1 + \rep{8_s} + \rep{27} + \rep{10} + \rep{10^{*}}
     + \rep{8_a}
\end{equation}
Among these six irreducible representations, first three,  $\rep1$,
$\rep{8_s}$, and
$\rep{27}$, are symmetric under the exchange of two baryons and the other three
$\rep{10}$, $\rep{10^{*}}$ and $\rep{8_a}$, are
antisymmetric.{\cite{DF}}
Noting that two-baryon states are to be
antisymmetric, and
that the color wave function for the color-singlet baryons is always symmetric,
we find that the symmetric (antisymmetric) flavor representations are combined
only to antisymmetric (symmetric) spin-orbital states.

In baryon-baryon scattering, the ALS force induces the transition
(mixing) between the spin singlet states ($^1P_1$, $^1D_2$, $^1F_3$, \ldots)
and the spin triplet states with the same $L$ and $J$
($^3P_1$, $^3D_2$, $^3F_3$, \ldots).
The flavor symmetries of $^1P_1$ state must be antisymmetric,
$\rep{10}$, $\rep{10^{*}}$ and $\rep{8_a}$, while
that of $^3P_1$ state is symmetric,  $\rep1$, $\rep{8_s}$ or $\rep{27}$.
If one assumes the \SU3\ invariance of the strong interaction,
different irreducible representations are not mixed.
Therefore the only possible combination of symmetric and antisymmetric
representations is $\rep{8_s}- \rep{8_a}$.
We conclude that the ALS in the SU(3) limit should only connect
$\rep{8_s}$ and $\rep{8_a}$.

The symmetry structure becomes clearer by decomposing
the $P$--wave $\Lam N- \Sig N$ ($I=1/2$), as a concrete example,
into the \SU3\ irreducible representations.
The flavor symmetric states read
\begin{equation}
  \pmatrix {\Lambda N \cr \Sigma N\cr} (^3P_1)
      =  \pmatrix {\sqrt{9/10} & -\sqrt{1/10} \cr
                   -\sqrt{1/10} &  -\sqrt{9/10} \cr }
        \pmatrix{ \rep{27} \cr \rep{8_s} \cr}
\end{equation}
while the antisymmetric ones are
\begin{equation}
  \pmatrix {\Lambda N \cr \Sigma N\cr}  (^1P_1)
      = \pmatrix {-\sqrt{1/2} & -\sqrt{1/2} \cr
                   -\sqrt{1/2} &  \sqrt{1/ 2} \cr }
        \pmatrix{ \rep{10*} \cr \rep{8_a} \cr}
\end{equation}
In the \SU3\ limit, the only  surviving matrix element is
$\langle \rep{8}_a \, ^1P_1 | V| \rep{8}_s \, ^3P_1\rangle$.
When we turn to the YN particle basis, we obtain the following
relations in the \SU3\ limit.
\begin{eqnarray}
  &&\langle \Lam N \,^1P_1 | V| \Sig N^{(1/2)}\, ^3P_1\rangle
  = - \, \langle \Sig N^{(1/2)} \,^1P_1 | V| \Sig N^{(1/2)}\,
  ^3P_1\rangle \nonumber\\
  &&\quad= 3\,\langle \Lam N \,^1P_1 | V| \Lam N\, ^3P_1\rangle
  = -3\, \langle \Sig N^{(1/2)} \,^1P_1 | V| \Lam N\, ^3P_1\rangle
\end{eqnarray}

On the contrary, the  $\Sigma N$ ($I=3/2$) system belongs purely to the
\SU3\ $\rep{27}$ and therefore the ALS matrix element vanishes in
the \SU3\ limit.
\begin{equation}
\langle \Sig N^{(3/2)} \,^1P_1 | V| \Sig N^{(3/2)} \,
^3P_1\rangle =0
\end{equation}

These relations come only from the \SU3\ symmetry and is general for
any ALS interactions regardless their origin.
We especially note that (1) the ALS for $\Sigma
N^{(1/2)}$ is much stronger than and has different sign from that
for $\Lam-N$,  (2) the coupling of $\Lam N - \Sig N^{(1/2)}$ is also strong,
and (3) the ALS for $\Sigma N$ depends strongly on the isospin or
the charge states.

\subsection{Quark cluster model approach}

In the quark cluster model approach,
the one-gluon exchange gives $q-q$ spin-orbit interaction, and
its contribution to the YN ALS forces is evaluated easily in the
adiabatic approximation.{\cite{Morimatsu}}
The results for the $P$ wave states at $R=0$,
where two baryons sit on top of each other,
are compared with the corresponding symmetric spin orbit (SLS) forces
in Table \ref{tab:QCM}.
The results show that the ALS forces due to the quark exchange
are as strong as the SLS force of the same origin.
Especially, the $\Sigma N$ ($I=1/2$) feels a stronger ALS
force between $S=0$ and $S=1$, than the SLS force
between $S=1$ states.
On the other hand,the ALS force vanishes in the $\Sig N$ ($I=3/2$),
such as $\Sig^+ p$ system.

\begin{table}[hbt]
\caption{ALS and LS matrix elements at $R=0$ normalized by the
overlapping matrix element.}
\label{tab:QCM}
\begin{tabular}{l|rr|rr}
\hline
&\multicolumn{2}{c|}{$\langle ^1P_1| V_{ALS}|^3P_1\rangle$}&
 \multicolumn{2}{c}{$\langle ^3P_1| V_{SLS}|^3P_1\rangle$} \\
&\SU3\ limit&\SU3\ broken&\SU3\ limit&\SU3\ broken\\
\hline
($I=1/2$)&(MeV)&(MeV)&(MeV)&(MeV)\\
$\Lam N \leftarrow \Lam N$& $37$ & $32$ & $-74$ & $-55$ \\
$\Lam N \leftarrow \Sig N$& $88$ & $77$ & $33$ & $29$ \\
$\Sig N \leftarrow \Lam N$& $-37$ & $-29$ & $33$ & $29$ \\
$\Sig N \leftarrow \Sig N$& $-88$ & $-79$ & $22$ & $22$ \\
\hline
($I=3/2$)&&&&\\
$\Sig N \leftarrow \Sig N$& $0$ & $1$ & $-95$ & $-94$\\
\hline
\end{tabular}
\end{table}

Table \ref{tab:QCM}\ also shows the potential values when the
\SU3\ symmetry is
broken by the mass difference of the strange quark and the $ud$ quarks.
The effects of the symmetry breaking are not so large that the results
are essentially the same.
Thus the above \SU3\ relations of the ALS matrix elements remain
valid qualitatively. (See ref.~\cite{TO} for details.)
This interesting result, that the quark exchange ALS force will
play dominant role in the YN forces, has been suggested and
demonstrated by Kyoto-Niigata group
in a realistic model of the YN interaction based on the quark
cluster model.{\cite{NSF}}

\subsection{Meson exchange force}

The ALS YN force in the meson exchange potential is known to be weak
compared to SLS.  The reason can again be understood qualitatively
from the \SU3\ symmetry. There the exchanged mesons
are either in the flavor singlet or octet representation
(because they are $q\bar q$ states).
The SU(3) factor for the (meson $M^a$)--(baryon $B_i$)--(baryon $B_j$)
coupling, $T^a_{ij}$, has three choices, $\delta_{ij}$
for the flavor singlet meson ($a=0$) and $F_{aij}$ or $D_{aij}$
for octet mesons ($a=1-8$),
where $F$ and $D$ are symmetric and antisymmetric SU(3)
structure constants, respectively.
Then the SU(3) invariant potential is proportional to
\[
 \sum_{a=0}^{8} \,( T^a_{ij} \cdot T^a_{lm} + \hbox{exchange term})
\]
One sees that the only possible antisymmetric coupling is of the form
$(F_{aij}\cdot D_{alm} - D_{aij}\cdot F_{alm})$.
This term, however, vanishes because the ratio of the
$F$ and $D$ couplings is fixed for each meson
without depending on the choice of
baryons ($ijlm$).
One exception is for the vector and the
tensor couplings in the vector meson exchange force.
According to the vector meson dominance, $F/D$ ratio for the vector
and the tensor couplings are in general different and then terms
like $(g_1 f_2 - f_1 g_2) (\vsig_1-\vsig_2)\cdot \vec L$
will survive, where $g_k$ ($f_k$) is the vector (tensor) coupling
constant of a vector meson to a baryon $k$ ($k=1$ or 2).
Thus the contributions of the meson exchanges to ALS YN force are
very limited in the \SU3\ limit, and we find that they are much
weaker than the SLS
counterparts.  The situation again does not change even if the
\SU3\ symmetry is broken.
See ref.~{\cite{TO}} for details.

Thus it is important to pin down the strengths and the properties
of the ALS forces in the YN sector as they indicate
the origin of the short-distance baryonic forces.
Further studies of the YN spin-orbit
interactions are very much encouraged.

\section{Direct Quark Process for the Weak Decay of $\Lambda$}

Recent experimental and theoretical studies of weak decays of
hypernuclei have generated renewed interest on nonleptonic
weak interactions of hadrons.
A long standing problem is the dominance of $\di=1/2$ amplitudes
in the strangeness changing transitions.
The decays of kaons, and $\Lam$, $\Sig$ hyperons are dominated
by the $\di=1/2$ transition but it is not clear whether this dominance
is a general property of all nonleptonic weak interactions.
In fact, the weak effective interaction which is derived
from the standard model including the perturbative QCD
corrections contains a significantly large $\di=3/2$
component.{\cite{hweak}}
It is therefore believed that nonperturbative QCD corrections,
such as hadron structures and reaction mechanism, are
responsible for suppression of $\di=3/2$, and/or enhancement
of $\di=1/2$ transition amplitudes.

From this viewpoint, decays of hyperons inside nuclear medium provide
us with a unique opportunity to study new types of nonleptonic
weak interaction, that is, two- (or multi-) baryon processes,
such as $\Lam N\to NN$, $\Sig N\to NN$, etc.
These transitions constitute the main branch of hypernuclear
weak decays because the pionic decay $\Lam\to N\pi$ is suppressed
due to the Pauli exclusion principle for the produced nucleon.

A conventional picture of the two-baryon decay process,
$\Lam N\to NN$, is the one-pion exchange between the baryons,
where $\Lam N\pi$ vertex is induced by the weak
interaction.{\cite{ope}}
In $\Lam N\to NN$, the relative momentum of the final state nucleon
is about 400 MeV/c, much higher than the nuclear Fermi momentum.
The nucleon-nucleon interaction at this momentum is dominated
by the short-range repulsion due to heavy meson exchanges and/or
to quark exchanges between the nucleons.
It is therefore expected that the short-distance interactions
will contribute to the two-body weak decay as well.
Exchanges of $K$, $\rho$, $\omega$, and $K^*$ mesons
and also correlated two pions in the nonmesonic
weak decays of hypernuclei have been
studied,{\cite{mesonex,Ramos}} and it is found that the kaon
exchange is significant, while the other mesons contribute
less.{\cite{Ramos}}

Several studies have been made on effects of quark
substructure.{\cite{CHK,MS,ITO}}
In our recent analyses,{\cite{ITO,IOMI}} we employ an effective weak
hamiltonian for quarks,
which takes into account one-loop perturbative QCD corrections
to the $W$ exchange diagram in the standard model.{\cite{hweak}}
We evaluated the effective hamiltonian in the six-quark wave
functions of the two baryon systems and derived the ``direct quark (DQ)''
weak transition potential for $\Lam N \to NN$.{\cite{ITO}}
Our analysis showed that the DQ contribution is significantly large
compared to the conventional pion exchange amplitudes, and shows some
qualitatively distinct features.
It largely improves the discrepancy between the meson-exchange theory
and experimental data for the ratio of the neutron- and proton-induced
decay rates of light hypernuclei.
It was also found that the $\di=3/2$ component of the effective hamiltonian
gives a sizable contribution to $J=0$ transition amplitudes.

\subsection{Nonmesonic Weak Decay of $\Lambda$ in Nuclei}

The DQ transition
takes place only when $\Lam$ overlaps with a
nucleon in hypernuclei and therefore predominantly in the relative
$S$-states of $\Lam N$ systems.
The two-body transition potentials in all the possible channels with
the initial $\Lam N (L=0)$ to the final $NN (L=0,1)$ states for the
DQ mechanism are computed.
Because of quark antisymmetrization effects, the DQ transition
potentials contain a nonlocal component and as the transition may
break the parity invariance, it also contains a derivative term.
The general form of the transition potential is
\begin{eqnarray}
 V^{\ell\ell'}_{ss'J}(r,r') &=& \langle NN: \ell' s' J|V({\vec r'},
  {\vec r}) |\Lam N : \ell s J\rangle \nonumber\\
  &=& V_{loc}(r)\, {\delta(r-r')\over r^{2}}
    + V_{der}(r)\, {\delta(r-r')\over r^{2}} \partial_{r}
    +V_{nonloc} (r',r)
\end{eqnarray}

We compare the DQ potential with conventional meson exchange
transition potentials, such as one-pion exchange (OPE).
Because the OPE potential is determined phenomenologically, the relation
between DQ and OPE is not trivial.  In order to fix the relative
phase of the two, we relate the $\pi N\Lam$
coupling constant to a baryon matrix element of the weak hamiltonian
for quarks by using a soft-pion relation,
\begin{equation}
\lim_{q\to 0}\langle\pi^{0}(q) n|H_{PV}|\Lam\rangle = {i\over f_{\pi}}
         \langle n|[Q_{5}^{3},H_{PV}]|\Lam\rangle = {-i\over 2f_{\pi}}
         \langle n| H_{PC} |\Lam\rangle
\end{equation}
Here we use the relation, $[Q_{5}^{3},H_{W}] = - [I^{3},H_{W}]$,
which is satisfied as the weak hamiltonian $H_{W}$ consists only
of left-hand currents and
the flavor-singlet right-hand currents.

In ref.~\cite{ITO,IOMI}, we calculated the nonmesonic decay rates of
$^{5}_{\Lam}He$, $^{4}_{\Lam}He$, and $^{4}_{\Lam}H$ in the DQ and
OPE mechanisms.
The $S$-shell hypernuclei are most suitable for the study of the
microscopic mechanism of the weak decay as their wave functions are
relatively simple and contain only $\Lam N (L=0)$ states.
They also enable us to select spin-isospin components for the weak
decay.
We found that DQ gives the major contribution in $J=0$ transitions and
therefore enhances the neutron induced decay rates.  The superposition
of the DQ and OPE shows a good agreement with available experimental
data.  We found that the $\Delta I=3/2$ components in the $J=0$
transitions are significant.
We also pointed out that the nonmesonic decay rate of
$^{4}_{\Lam}H$ is strongly enhanced by DQ, and therefore its
experimental data are critically important to confirm the DQ mechanism
for the nonmesonic weak decay of $\Lam$ in nuclei.

We here present results of the calculation of the $\Lambda$ decay in
nuclear matter.{\cite{SIO}}  We assume the $p-n$ symmetric nuclear matter with
realistic short-range correlation of $\Lam$ and $N$.
The results are summarized in Table 2.
We compare the results of several different combinations of the meson
exchanges and the direct quark processes.

\begin{table}[hbt]

\caption{Nonmesonic decay rates of ${\Lambda}$
in nuclear matter (in units of $\Gamma_{\Lambda}$).
The form factors are taken into account for the meson exchanges, where
the ``hard'' pion has $\Lam_{\pi}=1300 $MeV and the ``soft'' pion
has  $\Lam_{\pi}=800 $MeV.
The ``all'' includes $\pi$, $K$, $\eta$, $\rho$, $\omega$, and
$K^*$ meson exchanges.}
\begin{center}
\begin{tabular}{||c||c|c|c|c|c||}
\hline
\hline
 & total & $\Gamma_p$ & $\Gamma_n$ & ${\Gamma_n}/{\Gamma_p}$ & $PV/PC$ \\
\hline
$\pi$(hard) & 2.575 & 2.354 & 0.221 & 0.094 & 0.337 \\
$\pi$(soft) & 1.796 & 1.653 & 0.143 & 0.086 & 0.279 \\
\hline
$\pi$(hard)$+K$ & 1.099 & 1.076 & 0.024 & 0.022 & 0.631 \\
$\pi$(soft)$+K$ & 0.666 & 0.645 & 0.021 & 0.032 & 0.638 \\
\hline
$\pi$(hard) +all & 0.928 & 0.731 & 0.196 & 0.268 & 0.369 \\
$\pi$(soft) +all & 0.608 & 0.444 & 0.164 & 0.370 & 0.255 \\
\hline
DQ                   & 0.418 & 0.202 & 0.216 & 1.071 & 6.759 \\
\hline
DQ+$\pi$(hard)      & 3.609 & 2.950 & 0.658 & 0.223 & 0.856 \\
DQ+$\pi$(soft)      & 2.661 & 2.147 & 0.514 & 0.239 & 0.902 \\
\hline
DQ+$\pi$(hard)$+K$        & 1.766 & 1.495 & 0.271 & 0.181 & 1.602 \\
DQ+$\pi$(soft)$+K$        & 1.164 & 0.962 & 0.202 & 0.210 & 2.000 \\
\hline
DQ+$\pi$(hard) +all         & 1.507 & 1.123 & 0.384 & 0.342 & 1.471 \\
DQ+$\pi$(soft) +all         & 1.020 & 0.734 & 0.286 & 0.390 & 1.584 \\
\hline\hline
\end{tabular}
\end{center}
\end{table}

We notice that the kaon exchange reduces the
proton induced decay rates to more than factor two.  This mainly comes
from the suppression of the tensor transition, $\Lam N: ^3S_{1} \to NN:
^3D_{1}$.
On the other hand, the neutron induced decays are much too small in the
$\pi$ and $\pi +K$ exchanges, which is the main cause of the small $n/p$
ratio.
The DQ mechanism, however, enhances the neutron induced
decays and therefore improves the $n/p$ ratio.
It is also shown that the parity violating decay is dominant in DQ
transition, where the main component is the transition, $\Lam N:
^3S_{1} \to NN: ^3P_{1}$.

One sees in Table 2 that the pion exchange contribution strongly
depends on the choice of the form factor.
Some previous work uses a hard form factor with the cut off
$\Lam_{\pi} = 1300$ MeV according to the Bonn potential.
But such form factor in general gives too
large OPE contribution especially in the tensor transition,
$\Lam N: ^3S_{1} \to NN: ^3D_{1}$.
We here employ a softer form factor, $\Lambda\approx 800$ MeV,
which is suggested by an analysis of the pion production in
$NN$ scattering.{\cite{Lee-Matsuyama}}
It seems that the softer form factor is more appropriate to
reproduce experimental values of the proton induced decay rates.

Contribution beyond $\pi$ and $K$ mesons, especially the vector
mesons, contain some ambiguities.  For instance, the values and
even the signs of the weak coupling constants are not determined
phenomenologically.  They require \SU6\ ansatz.   They also
depend strongly on the form factors, which are not well known.
It is also questionable whether the DQ mechanism and the vector meson
exchanges are independent and can be superposed.
Here we follow the prescriptions given in ref.~\cite{Ramos}\ for
the vector exchange potentials and assume that there is no double
counting in superposing DQ with the vector meson exchanges.
We find a large $K^*$ contribution, which enhances the neutron
induced decay rate and thus improves the $n/p$ ratio.
We, however, do not think that this is the ``final'' result for the
vector meson contribution because of the above mentioned ambiguities
and unknown factors.

The reversed process, $pn\to\Lam p$, which is the hyperon weak
production in the $pn$ scattering, is also very interesting.
We calculated the cross section of the $\Lam$ production in the quark
cluster model, taking the full six-quark wave function into account.
The results will be published elsewhere.{\cite{ISO}}

\section{$\pi^+$ Decay of Hypernuclei}

In this section, we study
low energy $\pip$ emission in hypernuclear weak decays.
We point out that the soft $\pip$ decay is directly related
to $\di=3/2$ part of nonmesonic weak decays according to the
soft pion theorem.

The $\pip$ emission from light hypernuclei, for instance,
$^4_{\Lambda}He$, has puzzled us for a long time.
Rather old experimental data suggest that the ratio of $\pip$ and
$\pi^-$ emission from $^4_{\Lambda}He$ is about 5\%.{\cite{pip-data}}
This small ratio is expected because the free $\Lam$ decays only into
$p\pi^-$ and $n\pi^0$.  The $\pip$ emission requires an assistance
of a proton, \ie, $\Lam+p\to n+n+\pip$.

Several microscopic mechanisms for the $\pip$ emission
have been considered in literatures.{\cite{DH,CG,GT}}
The most natural one is $\Lam\to n\pi^0$ decay followed by
$\pi^0 p \to \pip n$ charge exchange reaction.
It was evaluated for realistic hypernuclear wave functions and
found to explain only 1.2\% for the $\pip/\pim$ ratio.{\cite{DH,CG}}
Another possibility is to consider $\Sig^+ \to \pip n$ decay
after the conversion $\Lam p \to \Sig^+ n$ by the strong interaction.
It was found, however, that the free $\Sig^+$ decay which is dominated
by $P$-wave amplitude, gives at most 0.2\% for the $\pip/\pim$ ratio.
Indeed it is clear that the $\Sig^+$ mixing and its free decay
is not the main mechanism, for experimental data suggest that
the $\pip$ emission is predominantly in the $S$-wave with the energy
less than 15 MeV.
Recently, it was proposed that a two-body process $\Sig^+ N\to n N\pip$
must be important in the $^4_{\Lambda}He$ decay.{\cite{GT}}
But its microscopic mechanism is not specified.

Here, we would like to show that the $\di=3/2$ two-baryon
transition amplitudes are directly related to the S-wave $\pip$
emission from hypernuclei.
The relation of these two amplitudes is derived from the soft-pion
theorem and is a result of the chiral structure of the weak interaction.

The soft-pion theorem for the process $\Lam p \to nn \pip(q\to 0)$
gives a similar relation to the one we use in the previous section,
\begin{equation}
  \lim_{q\to 0} \langle nn\pip(q)|H_W|\Lam p\rangle
= {i\over \sqrt{2} f_{\pi}} \langle nn|[I^-, H_W]|\Lam p \rangle
\label{soft-pion}
\end{equation}
As $H_W$ changes the third component of the isospin by $-1/2$
when it converts
$\Lam p$ ($I_3 = +1/2$) to $nn\pip$ ($I_3 = 0$),
it may contain $H_W(\di=1/2, \di_z=-1/2)$ and
$H_W(\di=3/2, \di_z=-1/2)$.
Now it is easy to see
that $\di=1/2$ part vanishes in eq.(9) as
\begin{eqnarray}
  [I_-,H_W(\di=1/2, \di_z=-1/2)] &=& 0 \\{}
  [I_-,H_W(\di=3/2, \di_z=-1/2)] &=& \sqrt{3} H_W(\di=3/2,
  \di_z=-3/2)
\end{eqnarray}
We then obtain
\begin{equation}
  \lim_{q\to 0} \langle nn\pip(q)|H_W|\Lam p\rangle
= {i\sqrt{3} \over \sqrt{2} f_{\pi}}
   \langle nn|H_W(\di=3/2, \di_z=-3/2)|\Lam p \rangle
   \label{eq:pipamp}
\end{equation}
Thus we conclude that the soft $\pip$ emission in the
$\Lam$ decay in hypernuclei is caused only by the $\di=3/2$
component of the strangeness changing weak hamiltonian.
In other words, the $\pip$ emission from hypernuclei probes the
$\di=3/2$ transition of $\Lam N \to NN$.

Now we understand why the previous attempts to explaining
the $\pip/\pim$ ratio failed.
Both the charge exchange process and the $\Sig^+$ decay are induced by
the $\di=1/2$ part of the hamiltonian and therefore cannot emit
low-energy $\pip$.
In fact, the reason why the $S$-wave $\Sig^+\to n\pi^+$ decay
is very small is again that only the $\di=3/2$ amplitude can induce
this decay for soft ($S$-wave) $\pi^+$.
In the same way, two-body $\Sigma^+$ decay will not contribute
unless it is induced by a $\di=3/2$ weak interaction.

\begin{table}[hbt]

\caption{Soft $\pip$ and $\pim$ decay rates in arbitrary units.}
\begin{center}
\begin{tabular}{||lr|r|r||}
\hline
 && DQ $+\pi +K$ & DQ $+\pi$ \\
\hline
$^4_{\Lam}\rm{He}$ &$\to \pim$  & 193.9 & 208.5  \\
 &$\to \pip$  & 65.4 & 65.4  \\
 &$\pip/\pim$  & 34\% & 31\%  \\
\hline
$^4_{\Lam}\rm{H}$ &$\to \pim$  & 169.1 & 211.1  \\
 &$\to \pip$  & 130.6 & 130.6 \\
 &$\pip/\pim$  & 77\% & 62\%  \\
\hline
$^5_{\Lam}\rm{He}$ &$\to \pim$  & 185.5 & 216.9  \\
 &$\to \pip$  & 65.4 & 65.4  \\
 &$\pip/\pim$  & 35\% & 30\%  \\
\hline
\end{tabular}
\end{center}
\end{table}

Supplying the $\Lam N\to NN$ two-body decay amplitudes given in the
previous section, we now
calculate the $\pip$ decay rates due to the two-body processes in the
soft pion limit.
Similar calculation can be done for the soft $\pim$ decay in the
two-body processes, where also the $\di=1/2$ components contribute.
Besides the direct quark process, the one-pion and one-kaon
exchanges are included in the $\di=1/2$ amplitudes,
while the $\di=3/2$ parts are purely from the direct quark mechanism.
The results are shown in Table 3.
We obtain the ratios of the $\pip$ emission to the
$\pim$ emission from the two-body processes are as large as
77\% in $^4_{\Lam}\rm{H}$, and 30-35 \% for $^4_{\Lam}He$ and $^5_{\Lam}He$.

In order to compare these ratios with experiment, we need to consider
the threshold differences of these decays carefully, as the phase
space volumes for the $\pip$ and $\pim$ decays are often largely
different.  We especially note, however, that the $^4_{\Lam}He$ decay
has a symmetric phase space volumes for $\pip$ and $\pim$ emissions
and therefore the $\pip$ decay branch is most easily observed.
In conclusion, it is extremely interesting to study the $\pip$ emission
carefully so that the $\Delta I=3/2$ component of the weak $\Lambda$
decay is confirmed.

\section{Conclusion}

We have exhibited several examples which represent the roles of
explicit quark content of the baryons in strangeness nuclear physics.
Both the strong and weak interactions in the YN systems show
characteristic features of the quark substructure.  It is in contrast
to the NN system, where most phenomena can be accounted either with
or without explicit quarks.  We hope that the YN system can
distinguish and enlighten the effects of quark substructure much more
clearly.  To this end, further efforts both in experimental and
theoretical studies are necessary.

\bigbreak
\noindent The authors thank Dr.\ S.~Takeuchi for useful discussions.
This work is supported in part by the Grant-in-Aid for scientific
research (C)(2)08640356 and Priority Areas (Strangeness Nuclear
Physics) of the Ministry of Education, Science and Culture of Japan.

\def \vol(#1,#2,#3){{{\bf {#1}} (19{#2}) {#3}}}
\def \NP(#1,#2,#3){Nucl.\ Phys.\          \vol(#1,#2,#3)}
\def \PL(#1,#2,#3){Phys.\ Lett.\          \vol(#1,#2,#3)}
\def \PRL(#1,#2,#3){Phys.\ Rev.\ Lett.\   \vol(#1,#2,#3)}
\def \PRp(#1,#2,#3){Phys.\ Rep.\          \vol(#1,#2,#3)}
\def \PR(#1,#2,#3){Phys.\ Rev.\           \vol(#1,#2,#3)}
\def \PTP(#1,#2,#3){Prog.\ Theor.\ Phys.\ \vol(#1,#2,#3)}
\def\MO{M.~Oka} \def\KY{K.~Yazaki}
\def \ibid(#1,#2,#3){{\it ibid.}\         \vol(#1,#2,#3)}
\def\MOka{\MO}
\def\KYazaki{\KY}
\def\Sachiko{\ST}
\def\PRD(#1,#2,#3){\PR(D#1,#2,#3)}
\def\PRC(#1,#2,#3){\PR(C#1,#2,#3)}
\def\etal{{\it et al.}}

\end{document}